\documentclass[11pt, titlepage]{article}

\usepackage{setspace}
\usepackage{amsmath}
\usepackage{amssymb}
\usepackage{graphicx}
\usepackage{color}
\usepackage[round, numbers, sort&compress]{natbib}
\def\tt{\texttt}
\newcommand{\mol}[1]{\left [ #1 \right ]}
\newcommand*{\unit}[1]{\ensuremath{\mathrm{\,#1}}}
\newcommand{\gate}[1]{\texttt{#1} gate}
\newcommand{\br}[1]{\left ( #1\right )}
\newcommand{\abs}[1]{\left\vert{#1}\right\vert}
\newcommand{\sqbr}[1]{\left [ #1\right ]}
\newcommand{\pdiff}[2]{\frac{\partial #1}{\partial #2}}
\newcommand{\avg}[1]{\langle{#1}\rangle}

\newcommand{\tref}[1]{Table \ref{table:#1}}

\newcommand{\tlabel}[1]{\label{table:#1}}
\newcommand{\fref}[1]{Fig.~\ref{fig:#1}}

\newcommand{\flabel}[1]{\label{fig:#1}}
\newcommand{\eref}[1]{Eq.~\ref{eqn:#1}}
\newcommand{\erefs}[1]{Eqs.~\ref{eqn:#1}}
\newcommand{\elabel}[1]{\label{eqn:#1}}

\title{Protein logic: a statistical mechanical study of signal integration at
the single-molecule level}
\author{Wiet~de~Ronde\thanks{FOM Institute AMOLF, Science Park 104, 1098 XG,
Amsterdam}\\
FOM Institute AMOLF\\
\and Pieter~Rein~ten~Wolde\\
FOM Institute AMOLF\\
\and Andrew~Mugler\\
FOM Institute AMOLF
}
\date{\today}

\begin{document}

\maketitle

\abstract{
Information processing and decision making is based upon logic operations,
which in cellular networks has been well characterized at the level of
transcription. In recent years however, both experimentalists and theorists have
begun to appreciate that cellular decision making can also be performed at the
level of a single protein, giving rise to the notion of protein logic.
Here we systematically explore protein logic using a well known statistical
mechanical model. As an example system, we focus on receptors which bind either
one or two ligands, and their associated dimers. Notably, we find that a single
heterodimer can realize any of the $16$ possible logic gates, including the
\tt{XOR} gate, by variation of biochemical parameters. We then introduce the
novel idea that a set of receptors with fixed parameters can encode functionally
unique logic gates simply by forming different dimeric combinations. An
exhaustive search reveals that the simplest set of receptors (two single-ligand
receptors and one double-ligand receptor) can realize several different groups
of three unique gates, a result for which the parametric analysis of single
receptors and dimers provides a clear interpretation. Both results underscore
the surprising functional freedom readily available to cells at the
single-protein level.
}

\clearpage
\section*{Introduction}

Cells depend on cues from their environment to initiate behaviors, including
growth, division, differentiation, and death. Based upon these environmental
signals cells must make decisions, such that the correct response is initiated.
Although a particular environmental signal often elicits a particular cellular
response, it is well established that signals can also act in combination
\citep{Mehta2009, Kaplan2008, Prehoda2002a, Bray1995}. In this case the
response triggered when two signals are present can be distinct from the
responses triggered by each signal alone. The cell thereby acts as a logic gate,
integrating two inputs to produce a single output. For example, the \gate{AND}
produces an output if both inputs are present, but it produces no output if
either a single input or no input is present. For the process of decision
making, the logic gate is the basic unit of computation, and therefore many
studies have been devoted to its role within biochemical networks.
Indeed, the role of logic gates within transcriptional networks has been
studied in depth: systematic theoretical studies have predicted
\citep{Hermsen2006, Bintu2005a, Istrail2005, Buchler2003} and experimental
studies have confirmed \citep{Tamsir2011,Kramer2004, Setty2003,	Guet2002} that
transcriptional networks can access all possible types of logic gate.

Recently, it has become clear that individual proteins can perform logic
operations as well. Although this notion was initially suggested
almost two decades ago \citep{Bray1995}, recent experiments have provided
striking demonstrations. For example, performance of an \gate{AND} by the
actin regulatory protein N-WASP has been observed {\em in vivo}
\citep{Prehoda2002a}:
when both of its inputs Cdc42 and PIP2 are present, they jointly ``unfold''
the active domain, leading to the activation of its target Arp2/3. Moreover,
synthetic proteins based upon naturally existing proteins have been constructed
and shown to perform a number of different logic operations
\citep{Dueber2004a,Dueber2003}.  Although these experiments beautifully
illustrate the capacity for single proteins to encode logic, they are restricted
to a limited set of logic gates. It therefore remains an open question if all
possible logic gates can be accessed by single proteins, in particular the
more complex gates like \tt{XOR}, which includes nonmonotonic behavior.

Despite the fact that transcriptional logic has been explored in depth, to our
knowledge no systematic theoretical study of protein logic has been done. A
recent study focused on the nonmonotonic behavior of a single protein with
several allosteric subunits, providing an understanding of how the action of a
ligand as an agonist or an antagonist could be switched by the presence of a
second ligand \citep{Motlagh2012}.
Beyond this nonmonotonic behavior, however, other mappings of ligand presence
to protein activity were not considered. By framing the problem as one of logic
computation, we here obtain a comprehensive understanding of the functional
mappings available to single proteins, thereby answering the question of which
logical functions are possible, and under what conditions. Moreover, we use a
less complex model than that used in \citep{Motlagh2012}, and we nonetheless
find rich functional behavior, including nonmonotonicity, as characterized by
the \gate{XOR}.

For several reasons, we focus on receptor proteins, although our approach is
easily extended to other protein types. First, receptors can be stimulated by
multiple ligands \citep{Citri2006a, Barton2002, Moghal1999, Bray1998}, which
naturally suggests a logic gate framework, in which multiple inputs (ligand
concentrations) are integrated into one output (receptor activity). Second,
receptors process signals directly at the plasma membrane. It is becoming
increasingly recognized that the plasma membrane is a hub of information
processing, acting as a mediator between the cell and its environment, along and
across which signals are stored, processed, and relayed \citep{Grecco2011}.
Receptors are integral to this process, as they affect decisions directly at
the detection level, before further intracellular transduction leads to the
ultimate cellular response. The encoding of logic by receptors thus has the
potential to be low-cost, since it is achieved with a single protein, and rapid,
since it occurs at the beginning of the signaling pathway. Finally, receptors
often exist in the form of dimers or higher oligomers. For example, G
protein-coupled receptors (GPCR) and ErbB receptors can each form dimers
consisting of receptors of the same type (homodimers) or of receptors of
different types (heterodimers) \citep{Landau2008, Minneman2007, Citri2006a}. A
dimer has the capacity to perform the same or more logic operations than each of
its monomeric constituents, a fact which we demonstrate here.
Moreover, as we describe for the first time here, dimerization permits function
space to be explored combinatorially: a cell can potentially change which
logical function is performed simply by modulating which combination of monomers
actually dimerizes.

We use a statistical mechanical model to develop a predictive framework for
protein logic. We first numerically probe the logic gates
accessible to individual receptor monomers and dimers by parameter variation,
which has relevance on evolutionary timescales. We find that
a single dimer can implement any of the possible logic gates with two inputs,
a result which we support analytically. Next, we introduce
the novel idea that a diverse set of logic gates can be performed, {\em not}
by variation of parameters, but by modulating the dimerization
of a fixed set of monomers. Such modulation may be achieved at the level of
transcription and translation of monomeric proteins, or via
post-translational modifications that enable monomers to dimerize; as such, we
argue that dimeric recombination provides a way for a cell to
modulate decisions on the timescales of gene expression or cell signaling. We
find that the simplest set of receptors (two single-ligand
receptors and one double-ligand receptor) can realize several different groups
of three unique gates and that together these groups include
all possible gates.
We provide clear analytic support for this result, following the previous
parametric analysis of single monomers and dimers. This result shines an
interesting new light onto why receptors, or proteins in general, exist in the
form of dimers. Both results underscore the surprisingly rich capacity for cells
to encode decisions using single molecules. Finally, throughout the study, we
discuss biological systems that implement these logical functions at the
single-protein level.

\section*{Methods}

We study receptor function by appealing to an equilibrium statistical
mechanical model. Statistical mechanical models have been used
quite fruitfully in the study of many molecular biology problems, including
receptor activity and gene regulation \citep{Phillips2008}.
In the case of receptors, several models are well known. All assume that a
receptor can exist in either an active ($A$) or an inactive
($I$) state, and that binding of a ligand changes the receptor bias for each
state. In the Koshland-Nemethy-Filmer (KNF) model, ligand
binding directly activates the receptor \citep{Koshland1966}. That is, the
bias is complete: a ligand-bound receptor is active, and an
unbound receptor is inactive. This condition is relaxed in the
Monod-Wyman-Changeux (MWC) model, in which ligand-bound receptors can be
in either state, but coupled receptors switch between states in synchrony
\citep{Monod1965}. Finally, in the conformational spread (CS)
model \citep{Changeux1967, Bray2004}, both conditions are relaxed: a
ligand-bound receptor can be in either state, and coupled receptors
can be in different states.
Because we are interested in the minimal model that can capture the ability to
perform logic gates, we adopt the MWC model; the KNF model
prohibits certain logic gates by construction, while the CS model allows
excess parametric freedom (clearly, what can be achieved by the MWC
model can be achieved by the CS model). Furthermore, the MWC model has been
shown to agree with experiments on receptors \citep{Tu2008,
Hansen2010,Skoge2006}.

The input in our model is the pair of concentrations $[S_1]$ and $[S_2]$ of
two different ligands. The output is the probability for a
receptor monomer or dimer to be in its active state. We consider three monomer
types and the associated dimers (\fref{setup}): a monomer
that binds ligand $1$ ($U$), one that binds ligand $2$ ($V$), and one that
binds both ligands ($W$). In the last case, ligand binding is
competitive: there is only one binding pocket, so only one ligand type can
bind at a time. Noncompetitive binding, in which both ligand
types can bind simultaneously, is captured by the $Q_{\rm UV}$ dimer
(\fref{setup}).

\begin{figure}[ht]
\begin{center}
\includegraphics[clip]{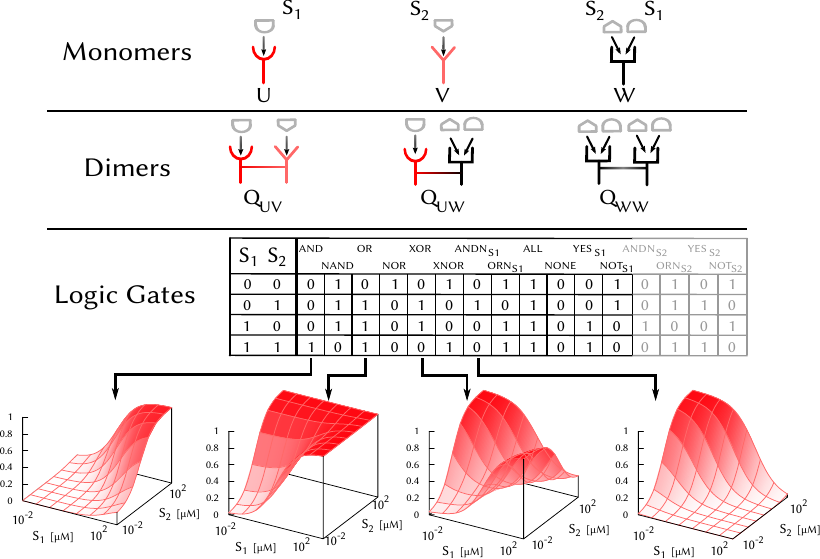}
\end{center}
\caption{\flabel{setup} Setup. We consider receptor monomers (top row) that bind
ligand $1$ (U),
ligand $2$ (V), or both competitively (W), and their associated
dimers (middle row). Bottom row: the table defines the $16$ possible two-input
logic gates in terms of binary input and output; below, for the four
functionally unique gates, we plot the continuous analogs given by the
statistical mechanical model.}
\end{figure}

The probability $p^A$ for a receptor to be in the active state is computed
from the partition functions, which enumerate all possible ways a
receptor can be in either the active ($Z^A$) or inactive ($Z^I$) state:
\begin{align}
\elabel{pA}
	p^A=\frac{Z^A}{Z^I+Z^A}.
\end{align}
The explicit forms of the partition functions under the MWC model are
presented as each monomer and dimer is discussed in the Results
section.
For intuition, we provide an example here: the partition functions for monomer
$W$ are

\begin{align}
	\elabel{ZA_W}
	Z^A&=\omega_0\br{1+\frac{\mol{S_1}}{K^A_1}+\frac{\mol{S_2}}{K^A_2}},\\
	\elabel{ZI_W}
	Z^I&=1+\frac{\mol{S_1}}{K^I_1}+\frac{\mol{S_2}}{K^I_2},
\end{align}
where the parameter $\omega_0=e^{-E_0/k_BT}$ is the Boltzmann factor
corresponding to the energy difference $E_0$ between the active and inactive
state, and the parameters $K_i^j$ are the dissociation constants of
ligand $i \in \{1,2\}$ in activity state $j \in \{A,I\}$. The variables
$\mol{S_1}$ and $\mol{S_2}$ are the ligand concentrations.
In \eref{ZA_W}, the three terms correspond to the
receptor being active when no ligand is bound, when
ligand $1$ is bound, and when ligand $2$ is bound, respectively. The same
holds for \eref{ZI_W} with the receptor being inactive.

The dependence of $p_A$ on  $\mol{S_1}$ and $\mol{S_2}$
defines the receptor's function (\fref{setup}, bottom row). Functions are
categorized based on the idealized behavior prescribed by the $16$ possible
two-input binary logic gates (\fref{setup}). Mathematically, the
function approaches binary logic when the output is either minimal ($p^A
\rightarrow 0$) or maximal ($p^A \rightarrow 1$) in each of the
four states defined by each input being absent ($\mol{S_1}=0$ or
$\mol{S_2}=0$) or present ($\mol{S_1}>0$ or $\mol{S_2}>0$).

Numerically, when varying parameters to assess whether a receptor
can realize a particular logic gate, we use a variant of the Wright-Fisher
algorithm \citep{Wright1931,Fisher1930}, which models the evolution of a
population.  In the Wright-Fisher algorithm, evolution
occurs in discrete, synchronous steps, and the population size
remains constant.  At each step, each member of the population produces
offspring in proportion to its fitness.  Then, mutations occur, and
the mutated offspring comprise the population for the next step.
In our case, for a given receptor, we have a ``population'' of $R$ initial
parameter points $\varphi_r$. Each point has fitness $f_r$, and the total
fitness for the receptor is $F=\sum_r f_r$. At each step, $R$ new points
(``offspring'') are drawn from the distribution $p_r = f_r/F$,
which weights each point by its fitness. Each new point is then ``mutated'' by
multiplying a randomly selected parameter by the factor $(1+\delta)$, where
$\delta$ is drawn uniform randomly from the range $[-\Delta:\Delta]$; we take
$\Delta=0.3$.

We define fitness as the agreement between the real-valued output of the
statistical mechanical model $p^A$ and the binary output of a
specific ideal logic gate. The ideal logic gate is prescribed by the goal
function $G\br{\mol{S_1}, \mol{S_2}}$,
which takes the value $0$ or $1$ depending on
whether each input is ``off''
($\mol{S_1} < \mol{S^*}$ or $\mol{S_2} < \mol{S^*}$)
or ``on''
($\mol{S_1} > \mol{S^*}$ or $\mol{S_2} > \mol{S^*}$),
where we take the threshold value $\mol{S^*} = 1 \unit{\mu M}$.
We compare $p^A$ and $G$ over an $N\times N$ grid of input values, spaced
logarithmically over the ranges of $\mol{S_1}$ and $\mol{S_2}$, which we take to
be $[10^{-2}-10^2] \unit{\mu M}$.
The fitness is thus
\begin{align}
\elabel{score}
f_r&=-\sum_{n,n' = 1}^{N}
\abs{p^A\br{[S^{n}_1],[S^{n'}_2]}-G\br{[S^{n}_1],[S^{n'}_2]}}.
\end{align}
The results in this article are obtained for $N=4$. Taking $N=2$ leads to
suboptimal results, while taking different values of $N>2$ yields
similar results to $N=4$.
Similar results are also obtained for a fitness function with $N=2$ and an
additional central point at $[S_1]=[S_2]=\mol{S^*}$,
at which $G$ is the average of the truth table for the gate.

The optimization parameters, as well as the bounds within which the model
parameters are initialized and constrained during
optimization, are given in Table SI-1 in the Supporting Material. The chosen
bounds fall within experimentally observed ranges and are consistent with
typical values used in previous modeling studies; we elaborate upon this point
in detail in Sec.\ SI.1.

When investigating whether multiple gates can be performed at fixed parameters
by formation of the possible dimer combinations, we optimize for several logic
gates at one time (Sec.\ SI.1).  In practice, a given point in parameter space
specifies both the dissociation constants $K_i^j$, which are intrinsic to each
monomer $U$, $V$, and $W$ and do not change when they are recombined, and the
Boltzmann factors $\omega_0$ and $\omega_{ii'}$, which are dimer-specific.

\section*{Results}

First, we identify the logic gates that each receptor monomer and dimer can
perform by parameter variation. Here, several derived analytic
constraints support the numerical results. Then, we investigate the extent to
which distinct logic gates can be formed using a set of
monomers with fixed parameters by forming the possible dimer combinations.
Several groups of distinct gates are possible, a finding for
which the first results provide a clear interpretation.

\subsection*{Functions accessible by parameter variation}

Figure~\ref{fig:param} shows the set of logic gates that each monomer and
dimer can perform, as determined by numerical optimization of
model parameters. The most striking feature is that one of the dimers can
perform all $16$ possible gates. This and the other
numerical results in \fref{param} can be understood intuitively by appealing
to analytic results derived from the underlying model, which we will describe in
turn for each monomer and dimer.

\subsubsection*{Monomers}

The first two monomers, receptors $U$ and $V$, respond to only one input each.
Therefore, they are trivially constrained to gates which
depend on neither input (\tt{ALL}, \tt{NONE}) or on only one input
(\tt{YES}$_{S_i}$, \tt{NOT}$_{S_i}$). Receptor $W$, on the other hand,
allows competitive binding of both inputs, and can therefore realize several
nontrivial gates.

At this point it is useful to observe that the gates exist in antagonistic
pairs (a gate and its inverse), shown consecutively in
\fref{param}: (\tt{AND}, \tt{NAND}), (\tt{OR}, \tt{NOR}), etc. Any receptor
that can perform one member of a pair can perform the other,
simply by inverting certain parameter values.
Furthermore, several gates are equivalent under reversal of the two ligands
(those with subscripts in \fref{param}):
(\tt{ANDN$_{S_1}$}, \tt{ANDN$_{S_2}$}), (\tt{ORN$_{S_1}$}, \tt{ORN$_{S_2}$}),
etc. Again, any receptor that can perform one of these can
perform the other, simply by switching certain parameter values (corresponding
to exchanging the effect of $S_1$ and $S_2$). Eliminating
these redundancies, we arrive at four unique gates that respond nontrivially
to both inputs:
\begin{align}
\elabel{gates}
\tt{AND}, \quad \tt{OR}, \quad \tt{ANDN}_{S_1}, \quad \tt{XOR}.
\end{align}
We will consider only these four unique gates from this point on.

The third monomer, receptor $W$, whose partition functions are given in
\erefs{ZA_W}-\ref{eqn:ZI_W}, can realize two of the four unique
gates: \tt{OR} and \tt{ANDN}$_{S_1}$. The \gate{OR} follows straightforwardly
from the situation where both ligands activate the receptor
individually; their combination will then activate it as well.
The \gate{ANDN$_{S_1}$} can be formed if ligand $1$ binds more strongly than
ligand $2$ ($K_1^j \ll K_2^j$), but ligand $1$ only weakly biases the receptor
toward the active state ($K_1^A \sim K_1^I$), while ligand
$2$ strongly biases it ($K_2^A \ll K_2^I$). In this scenario, a receptor that
is inactive in the absence of both ligands ($\omega_0 \ll 1$)
will only be active in the presence of ligand $2$ {\it and not} $1$.

We note here that any receptor that is activated by two different ligands is a
biological example of an \tt{OR} gate, and many naturally
occurring receptors are activated by different ligands, like the TAR receptor
\citep{Bray1998} and the EGF receptor \citep{Jorissen2003}.
More generally, it has been shown that proteins can be synthesized with a
number of specific ligand-binding sites \citep{Looger2003}; such constructs can
be thought of as extensions of the \gate{OR} to more than two inputs.
Additionally, the \gate{ORNOT}, the inverse of the \gate{ANDN}, has been
constructed synthetically using a single protein (construct H2, Fig. 2B in
\citep{Dueber2003}).

Receptor $W$ cannot realize the other two unique gates, \tt{AND} and \tt{XOR}.
Both gates require a cooperative effect when both ligands
are present: in the \gate{AND}, neither ligand activates the receptor
individually, but both activate it together; in the \gate{XOR}, each
ligand activates the receptor individually, but both suppress activation
together. Such cooperative effects are not possible with
competitive binding. As we will see next, dimerization is required to perform
these gates.

\subsubsection*{Dimers}

The three monomers admit six possible dimer combinations --- three homodimers
and three heterodimers. The homodimers $Q_{\rm UU}$ and $Q_{\rm VV}$ respond to
only one input each and are therefore trivially constrained like monomers $U$
and $V$. Moreover, heterodimers $Q_{\rm UW}$ and $Q_{\rm WV}$ are equivalent
upon ligand exchange and can therefore realize equivalent sets of logic gates
upon parameter variation. This leaves three dimers that can realize unique sets
of logic gates upon parameter variation: $Q_{\rm UV}$, $Q_{\rm UW}$, and
$Q_{\rm WW}$.

The first dimer, receptor $Q_{\rm UV}$, is the simplest heterodimer: it is
formed by combining monomer $U$, which responds only to ligand
$1$, and monomer $V$, which responds only to ligand $2$. Unlike receptor $W$,
which is limited to competitive binding, the dimeric receptor
$Q_{\rm UV}$ has two binding pockets and therefore allows noncompetitive (i.e.\
cooperative) binding. Accordingly, its partition functions extend
those of receptor $W$ (\erefs{ZA_W}-\ref{eqn:ZI_W}) to include a cooperative
term:

\begin{align}
\elabel{ZA_QUV}
Z^A&=\omega_0
\br{1+\frac{\mol{S_1}}{K^A_1}+\frac{\mol{S_2}}{K^A_2}+\overbrace{\omega_{12}
\frac { \mol{S_1}}{K^A_1} \frac{\mol{S_2}}{K^A_2}}^ {\rm
cooperative}},\\
\elabel{ZI_QUV}
Z^I&=1+\frac{\mol{S_1}}{K^I_1}+\frac{\mol{S_2}}{K^I_2}+\omega_{12}\frac{\mol{S_1
}}{K^I_1}\frac{\mol{S_2}}{K^I_2},
\end{align}
The cooperative term contains an additional Boltzmann factor
$\omega_{12}=e^{-E_{12}/k_BT}$ corresponding to the cooperative binding energy
$E_{12}$, which could originate from, e.g., a conformational change of the
receptor upon binding of one ligand that opens the binding pocket for the other
ligand.
For example, the binding affinity of each of the inputs to the protein N-WASP
is increased by a factor of $\sim$$300$ when the other input is bound
\citep{Prehoda2002a}.

Receptor $Q_{\rm UV}$ can realize three of the four unique gates: \tt{OR},
\tt{ANDN}$_{S_1}$, and \tt{AND}. The \tt{OR} and
\tt{ANDN}$_{S_1}$ gates follow straightforwardly from the fact that $Q_{\rm
UV}$ reduces to $W$ for no cooperativity ($\omega_{12}=0$), and
receptor $W$ can realize these gates as previously discussed.
The \gate{AND} is formed when the receptor is inactive in the presence of each
ligand alone but, due to the cooperative interaction, is
active in the presence of both ligands together.
Receptor $Q_{\rm UV}$ cannot realize the \gate{XOR}: if the receptor is
activated by either one of the two ligands, it must also be
activated by both ligands together. The cooperative interaction enhances the
effect that each ligand individually has on the activation of
the receptor, but it cannot reverse it.

The intuition behind why receptor $Q_{\rm UV}$ can realize the \gate{AND} can
be quantified by considering the constraints that an
\gate{AND} places on the partition functions:

\begin{align}
\elabel{AND_constraints}
\begin{tabular}{rrr|l}
$\mol{S_1}$ & $\mol{S_2}$ & $p^A$ &\\
\hline
$0$ & $0$ & $0$ & $\,\, \omega_0\ll 1$\\
$\bar{\mol{S_1}}$ & $0$ & $0$ &
	$\,\, \omega_0\br{1+\frac{\bar{\mol{S_1}}}{K^A_{1}}}\ll
1+\frac{\bar{\mol{S_1}}}{K^I_{1}}$\\
$0$ & $\bar{\mol{S_2}}$ & $0$ &
	$\,\, \omega_0\br{1+\frac{\bar{\mol{S_2}}}{K^A_{2}}}\ll
1+\frac{\bar{\mol{S_2}}}{K^I_{2}}$\\
$\bar{\mol{S_1}}$ & $\bar{\mol{S_2}}$ & $1$ &
	$\,\,
\omega_0\br{1+\frac{\bar{\mol{S_1}}}{K^A_{1}}+\frac{\bar{\mol{S_2}}}{K^A_{2}}
	+\omega_{12}\frac{\bar{\mol{S_1}}\bar{\mol{S_2}}}{K^A_{1}K^A_{2}}}\gg
	1+\frac{\bar{\mol{S_1}}}{K^I_{1}}+\frac{\bar{\mol{S_2}}}{K^I_{2}}
	+\omega_{12}\frac{\bar{\mol{S_1}}\bar{\mol{S_2}}}{K^I_{1}K^I_{2}}$
\end{tabular}
\end{align}
Here, $\bar{\mol{S_1}}$ and $\bar{\mol{S_2}}$ denote the maximum input values.
We have recognized that a low output
requires $Z^A \ll Z^I$ (see \eref{pA}); therefore, the first three lines
reflect that in an \gate{AND} the output is low in the first three
input conditions. Similarly, a high output
requires $Z^A \gg Z^I$, which is reflected in the last line.
Receptor $Q_{\rm UV}$ can realize the \gate{AND} precisely because the
constraints in \eref{AND_constraints} can be met simultaneously. For
example, taking for illustration the simplifying case of intermediate
cooperativity ($\omega_{12} \gtrsim 1$) and symmetric, saturating
ligand concentrations
($\bar{\mol{S_1}}/K^j_1 = \bar{\mol{S_2}}/K^j_2 \gg1$),
\eref{AND_constraints} reduces to

\begin{align}
\elabel{AND_constraints_simple}
1 \ll 1/\sqrt{\omega_0} \ll K^{I}_{1}/K^A_{1} \ll 1/\omega_0.
\end{align}

Indeed, we see that the \gate{AND} requires a bias upon ligand binding that is
too weak to activate the receptor individually
($K^{I}_{1}/K^A_{1} \ll 1/\omega_0$), but strong enough to activate the
receptor cooperatively ($K^{I}_{1}/K^A_{1} \gg 1/\sqrt{\omega_0}$).

The strength of the cooperativity influences the quantitative properties of
the
\gate{AND}: an increase in $\omega_{12}$ shifts the transition region of the
gate to smaller ligand concentrations, as indeed observed in
studies of the \tt{AND}-like N-WASP protein \citep{Prehoda2002a}.

In addition to N-WASP \citep{Prehoda2002a}, the \gate{AND} logic is
observed in various other proteins. For example, in gonadotropes, the scaffold
PEA-15 is activated only by the simultaneous presence of PKC and ERK
\citep{Choi2011}.  Similarly, the adaptor protein TIRAP functions as a
coincidence detector \citep{Pawson2007}, thereby only becoming activated when
two inputs are present at the same time. In {\em V. Harveyi},
coincidence detection is also exhibited for the two quorum signals AI-1 and AI-2
\cite{Henke2004}.

The second dimer, receptor $Q_{\rm UW}$, is also a heterodimer: it is formed
by combining monomer $U$, which responds only to ligand $1$,
and monomer $W$, which responds competitively to both ligands.
The partition functions for this receptor are

\begin{align}
\elabel{ZA_QUW}
Z^A&=\omega_0 \Big(
1+\frac{\mol{S_1}}{K^A_{1,U}}+\frac{\mol{S_1}}{K^A_{1,W}}+\frac{\mol{S_2}}{K^A_{
2}}\nonumber\\
&\quad\quad\quad+\omega_{11} \frac{\mol{S_1}^2}{K^A_{1,U}K^A_{1,W}}
+\omega_{12}\frac{\mol{S_1}\mol{S_2}}{K^A_{1,U}K^A_{2}}
\Big),\\
\elabel{ZI_QUW}
Z^I&=1+\frac{\mol{S_1}}{K^I_{1,U}}+\frac{\mol{S_1}}{K^I_{1,W}}+\frac{\mol{S_2}}{
K^I_{2}} \nonumber\\
&\quad\quad\quad+\omega_{11}\frac{\mol{S_1}^2}{K^I_{1,U}K^I_{1,W}}+\omega_{12}
\frac{\mol{S_1}\mol{S_2}}{K^I_{1,U}K^I_{2}}.
\end{align}
Here, since ligand $1$ can bind to either monomer $U$ or $W$, we distinguish
these cases with the second subscript on $K^j_1$. There are
now two cooperative terms, corresponding to the cases where monomers $U$ and
$W$ bind, respectively, ligands $1$ and $1$ ($\omega_{11}$), or
ligands $1$ and $2$ ($\omega_{12}$).
\erefs{ZA_QUW}-\ref{eqn:ZI_QUW} make clear that receptor $Q_{\rm UW}$ reduces
to receptor $W$ (\erefs{ZA_W}-\ref{eqn:ZI_W})
in the limit $K^j_{1,U}\to\infty$, and to receptor $Q_{\rm UV}$
(\erefs{ZA_QUV}-\ref{eqn:ZI_QUV}) in the limit
$K^j_{1,W}\to\infty$.

Receptor $Q_{\rm UW}$ can realize all four unique gates (and therefore all
$16$ possible gates; \fref{param}). The \tt{OR},
\tt{ANDN}$_{S_1}$, and \tt{AND} gates follow straightforwardly from the fact
that receptor $Q_{\rm UV}$, which can realize these gates, is a
limiting case.
The \gate{XOR} is less trivial. Below we offer an intuitive argument for why
receptor $Q_{\rm UW}$ can realize an \gate{XOR}, and in
Sec.~SI.2 we
prove analytically that the output can be a nonmonotonic function of the two
inputs for this receptor, which is required for an \gate{XOR}.

The \gate{XOR} is formed when each ligand individually activates the receptor
by binding to monomer $W$, but when both ligands are present,
ligand $1$ is outcompeted and thus binds to monomer $U$, in turn suppressing
activation.
It is instructive here to describe this process in more detail. Suppose that
ligand $1$ promotes activation when bound to $W$ but
suppresses activation when bound to $U$. Further, suppose that ligand $1$
binds more strongly to $W$ than to $U$, such that in the presence
of ligand $1$ alone, the receptor is active. Now suppose that ligand $2$
promotes activation when bound to $W$. Since ligand $2$ can only
bind to $W$, in the presence of ligand $2$ alone, the receptor is also active.
Finally, suppose that ligand $2$ ``interferes'' with ligand
$1$, i.e.\ binds more strongly to $W$ than ligand $1$ does. Then, in the
presence of both ligands, ligand $2$ binds to $W$, leaving ligand
$1$ to bind to $U$. If $U$ suppresses activation more strongly than $W$
promotes activation, then in the presence of both ligands, the
receptor is inactive. The resulting logic is the \gate{XOR}.

The third dimer, receptor $Q_{\rm WW}$, is a homodimer: it is formed by
combining two $W$ monomers, each of which responds competitively to
both ligands.
The partition functions for this receptor are

\begin{align}
\elabel{ZA_QWW_homo}
Z^A&=\omega_0 \Big( 1+2\frac{\mol{S_1}}{K^A_{1}}+2\frac{\mol{S_2}}{K^A_{2}}
\nonumber\\
&\quad\quad\quad+\omega_{11}\frac{\mol{S_1}^2}{K^A_{1}K^A_{1}}+2\omega_{12}
\frac{\mol { S_1}\mol{S_2}}{
K^A_{1}K^A_{2}}+\omega_{22}\frac{\mol{S_2}^2}{K^A_{2}K^A_{2}} \Big), \\
\elabel{ZI_QWW_homo}
Z^I&=1+2\frac{\mol{S_1}}{K^I_{1}}+2\frac{\mol{S_2}}{K^I_{2}} \nonumber\\
&\quad\quad\quad+\omega_{11}\frac{\mol{S_1}^2}{K^I_{1}K^I_{1}}+2\omega_{12}
\frac{\mol { S_1}\mol{S_2}}{K^I_{1}K^I_{2}}
+\omega_{22}\frac{\mol{S_2}^2}{K^I_{2}K^I_{2}}.
\end{align}
Here, the factors of two account for the fact that each ligand can be bound to
either of two symmetric monomers. There are now three
cooperative terms, corresponding to the cases where both monomers bind ligand
$1$ ($\omega_{11}$), both bind ligand $2$ ($\omega_{22}$), or
one binds ligand $1$ and the other binds ligand $2$ ($\omega_{12}$).

Receptor $Q_{\rm WW}$ can realize three of the four unique gates: \tt{OR},
\tt{ANDN}$_{S_1}$, and \tt{AND}. The \tt{OR} and
\tt{ANDN}$_{S_1}$ gates follow straightforwardly from the fact that each
monomer alone can realize these gates as previously discussed. The
\gate{AND} relies on strong suppression of cooperation between monomers if
they are bound to the same ligand type (i.e.\ $\omega_{11}
\rightarrow 0, \omega_{22} \rightarrow 0$); this suppression prevents
activation when only one ligand is present.
In fact, this limit reduces \erefs{ZA_QWW_homo}-\ref{eqn:ZI_QWW_homo} to
\erefs{ZA_QUV}-\ref{eqn:ZI_QUV} (up to factors of $2$),
meaning the \gate{AND} constraint, \eref{AND_constraints_simple}, also holds
here under the same conditions for which it was derived.
Receptor $Q_{\rm WW}$ cannot realize the \gate{XOR}: because the individual
monomers are identical, no negative interference is possible, as
it is for receptor $Q_{\rm UW}$.

\begin{figure}[ht]
\begin{center}
\includegraphics[clip]{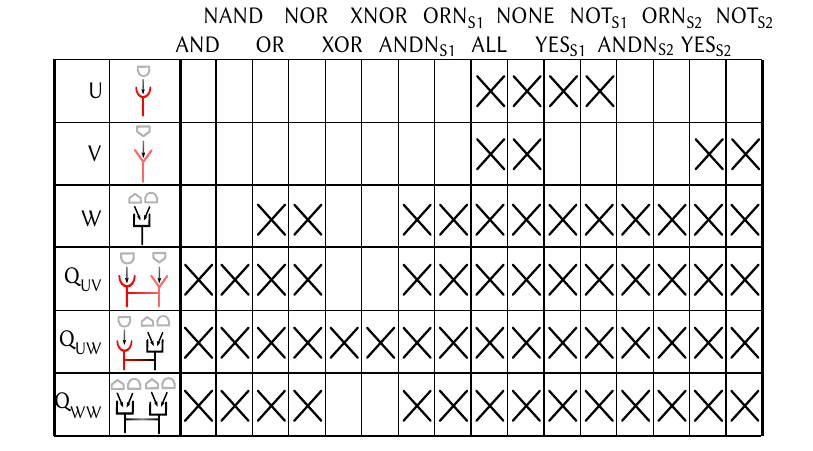}
\end{center}
\caption{\flabel{param} Functional versatility by parameter variation. For all
monomers and dimers, we
show the possible functions attainable by varying
parameters. Attainability is assessed by numerical optimization and
interpreted based on analytic constraints derived in the text.}
\end{figure}

\subsection*{Functions accessible by recombination}

In the previous section we identified the logic gates accessible by individual
receptors via variation of intrinsic biochemical parameters. In this section we
ask a separate question.
We here seek the logic gates that a set of monomers can realize --- at
fixed parameters --- simply by forming the possible dimer combinations.
This question is critically related to the challenge that all cells face:
to encode reliable responses
using limited resources (here, a limited set of monomers) and on short
timescales (here, set by gene expression and cell signaling).
This question is also key to functional control at the single-protein level: if
diverse logic gates can be realized by a small set of monomers,
cellular function could be strongly tuned in a straightforward
manner, e.g.\ by expressing a particular pair of monomers and not others.

The three monomers we study form four functional dimers: $Q_{\rm UV}$, $Q_{\rm
UW}$, $Q_{\rm WV}$, and $Q_{\rm WW}$ (the dimers $Q_{\rm UU}$
and $Q_{\rm VV}$ respond to only one input each and are neglected).
This fact leads to the
enticing question of whether there exist parameters at which the
four dimers perform the four unique logic gates (\eref{gates}). Such a finding
would be highly nontrivial: all monomers are present in at
least two dimers, and therefore the performance of a particular logic gate by
one dimer places heavy constraints on the parameters of the
other dimers.

An exhaustive search, in which we numerically optimize for each of the $4!=24$
dimer-to-logic gate mappings in turn (Sec.~SI.1),
suggests that no parameter set exists at which all four unique logic gates are
performed. Moreover, replacing any subset of gates with the
corresponding inverse gates and repeating gives the same result in each of the
$2^4=16$ cases. Interestingly, the result seems to be due to
the fact that the parameters which support the \gate{XOR} in receptor $Q_{\rm
UW}$ (or its counterpart $Q_{\rm WV}$) prohibit the \gate{AND}
in any of the other receptors. Next we support this numerical observation with
an intuitive argument.

Suppose that receptor $Q_{\rm UW}$ performs an \gate{XOR}.
As described in the previous section, the \gate{XOR} requires that when ligand
$1$ is present alone, it activates the receptor by binding to
monomer $W$. Since the \gate{AND} requires the opposite behavior, namely that
the receptor is inactive when ligand $1$ is present alone,
then the \gate{AND} cannot be formed by any receptor in which ligand $1$ only
binds to $W$. This group includes receptors $Q_{\rm WW}$ and
$Q_{\rm WV}$, leaving only receptor $Q_{\rm UV}$. Then, as also described in
the previous section, the \gate{XOR} requires that ligand $1$
suppresses activation when bound to monomer $U$. Since the \gate{AND} requires
that the receptor is active when both ligands are present,
in receptor $Q_{\rm UV}$ this suppression would have to be overpowered by
activation via ligand $2$ binding to $V$. However, if this were
the case, the receptor would surely be active in the presence of ligand $2$
alone, which is inconsistent with the behavior of an \gate{AND}.
These arguments make clear that if receptor $Q_{\rm UW}$ performs an
\gate{XOR}, no other receptor can form an \gate{AND}. The same
arguments, but with the ligands exchanged, hold if receptor $Q_{\rm WV}$
performs the \gate{XOR} instead of receptor $Q_{\rm UW}$. Since
receptors $Q_{\rm UW}$ and $Q_{\rm WV}$ are the only receptors that can
perform the \gate{XOR}, we conclude that the \tt{XOR} and \tt{AND}
gates are not mutually accessible by recombination of monomers $U$, $V$, and
$W$ at fixed parameters.

Even though all four unique logic gates cannot be performed at fixed
parameters, we do find six parameter sets at which unique groups of
three logic gates are performed by three of the dimers. We denote these
parameter sets as $\varphi_k$, for $k \in \{1, 2, \dots, 6\}$, and
show the logic gates and the dimers that perform them in
\fref{recombination}. We stress that this result is still
nontrivial: two of the groups are performed by receptors $Q_{\rm WW}$, $Q_{\rm
UW}$, and $Q_{\rm WV}$, which all contain monomer $W$;
additionally,
two groups are performed by receptors $Q_{\rm UW}$, $Q_{\rm WV}$, and $Q_{\rm
UV}$, in which each monomer is represented in two of
the three dimers. Due to the high degree of monomer overlap in both cases, one
might have expected the three dimers to be constrained to
similar functionality at fixed parameters; instead, we find that three unique
logic gates can be formed. Further, \fref{recombination}
reveals that all four logic gates are represented among the six groups (but,
as expected, never \tt{XOR} and \tt{AND} in the same group).
Finally, the optimal solutions shown in \fref{recombination} are robust to
parametric perturbation: as shown in Sec.~SI.3, for all
$\varphi_k$, most random perturbations in which each parameter changes by an
average of $\sim$$20\%$ change the fitness of none of the
three logic gates by more than $10\%$.
All of these features underscore the functional versatility available to cells
by dimeric recombination.

Our finding that cells can perform multiple logic gates at fixed
parameters naturally raises the question of whether the gates conflict with each
other, which could potentially corrupt the computation. 
Moreover, since we imagine that the
dimers are present on the membrane in quasi-equilibrium with their monomeric
constituents, we must also consider whether the gates are in conflict with the
logic encoded by the monomers themselves.  This latter question is
straightforward to resolve.  First, the monomers $U$ and $V$ respond to only one
input each and therefore do not perform nontrivial logic gates.  Second,
while the monomer $W$ can perform one of two nontrivial logic gates (\tt{OR} or
\tt{ANDN}), the dimer $Q_{\rm WW}$ then also performs this gate.  Any conflict
between $W$ and a dimer therefore also arises as a conflict between $Q_{\rm WW}$
and that dimer.  We thus consider only conflict between dimers from here on.

One simple way of minimizing conflict between dimers is by selectively
expressing only a particular pair of monomers and not the other monomer (\fref{conflict}a).  For
example, at parameter set $\varphi_3$ (see \fref{recombination}), if monomers $U$
and $V$ were expressed, but not $W$, the only dimer that could form is
$Q_{\rm UV}$, resulting in the unambiguous encoding of an \gate{ANDN}.  If at
some later time, monomers $U$ and $W$ were expressed, but not $V$, only $Q_{\rm
UW}$ and $Q_{\rm WW}$ could form; then, since $Q_{\rm WW}$ is not functional at
$\varphi_3$, the \gate{XOR} would be encoded unambiguously.  Similarly,
expression of $V$ and $W$ but not $U$ would encode the \gate{OR}
unambiguously.  The time between these periods of selective expression would be
set by gene expression and would therefore be long compared to the timescale on
which the cell actually employs the logic gate to respond to the incoming
signals.   We observe from \fref{recombination} that both parameter sets
$\varphi_3$ and $\varphi_4$
share the property that all three gates can be encoded unambiguously by
selective expression; in this sense they are ``optimal'' in terms of minimizing
conflict between gates.  By contrast, the other parameter sets suffer from
conflict between $Q_{\rm UW}$ and $Q_{\rm WW}$ when only $U$ and $W$ are
expressed ($\varphi_1$, $\varphi_2$, $\varphi_5$, $\varphi_6$), or between
$Q_{\rm VW}$ and $Q_{\rm WW}$ when only $V$ and $W$ are expressed ($\varphi_1$,
$\varphi_2$).

Ultimately, the most general solution to the problem of dimer conflict --- and
indeed, one that is commonly exploited by cells --- is to make the downstream
response dimer-specific.  Specificity can be established in several ways.  The
immediate downstream component can respond preferentially to one dimer and not to
another, as observed for the EGF receptor family \cite{Sweeney2001}.  The
specificity could then be propagated further downstream, for example at the
level of transcriptional regulation (\fref{conflict}b).
Alternatively, specificity can be achieved via spatial segregation of membrane
components (\fref{conflict}c).  For example, interaction with lipid rafts is
thought to separate membrane proteins into spatially distinct, non-mixed
clusters, leading to added specificity in downstream computations
\cite{Lingwood2010, Kholodenko2010}.  Either of these
mechanisms would allow several types of dimers to coexist on the membrane and
control, simultaneously and without conflict,  distinct downstream processes
according to distinct logical functions.

In the remainder of this section, we provide for parameter sets $\varphi_1$
and $\varphi_2$ the intuition behind how the three logic gates
in \fref{recombination} are performed by the corresponding receptors. In
Sec.~SI.4, we provide similar intuition for parameter sets $\varphi_3$,
$\varphi_4$, $\varphi_5$, and $\varphi_6$. Furthermore, in Sec.~SI.4, we argue
why the groups observed in \fref{recombination} (and their counterparts
obtained upon ligand exchange) are the {\it only} groups of three unique logic
gates that one expects to observe under this model.

Parameter set $\varphi_1$ (\fref{recombination}, first row) corresponds to a
case where each ligand only weakly promotes activation in the
presence of monomer $W$. This feature allows receptor $Q_{\rm WW}$ to remain
inactive when each ligand is present individually but become
activated when both ligands are present together, forming the \gate{AND}.
Furthermore, ligand-bound $U$ both promotes activation and
strongly enhances the binding of ligand $2$ to $W$. This feature allows
receptor $Q_{\rm UW}$ to perform the \gate{OR}: when ligand $1$ is
present alone, it promotes activation by binding to $U$; when ligand $2$ is
present in abundance and ligand $1$ is present only in a small
amount (and thus still in the ``off'' state), the small amount of ligand $1$
is nonetheless sufficient to
promote activation via enhanced binding of ligand $2$ to $W$; and when both
ligands are present in abundance, the two effects combine,
resulting in activation. Finally, ligand-bound $V$ both suppresses activation
and strongly enhances the binding of
ligand $1$ to $W$. This feature allows receptor $Q_{\rm WV}$ to perform the
\tt{ANDN}$_{S_2}$ gate: when ligand $2$ is present it
suppresses activation via $V$, independent of ligand $1$; but when ligand $1$
and not (very much of) ligand $2$ is present, the small amount
of ligand $2$ strongly enhances binding of ligand $1$ to $W$, thus promoting
activation.

Parameter set $\varphi_2$ (\fref{recombination}, second row) corresponds to a
case where ligand-bound $W$ promotes activation. This
feature is sufficient for receptor $Q_{\rm WW}$ to perform the \gate{OR}.
Furthermore, ligand $2$ binds more strongly to $V$ than to $W$,
and ligand-bound $V$ suppresses activation more strongly than ligand-bound $W$
promotes activation. These features allow receptor $Q_{\rm
WV}$ to perform the \tt{ANDN}$_{S_2}$ gate, since only in the presence of
ligand $1$ and not $2$ will activation be promoted via $W$ and not
suppressed via $V$. Finally, (i) ligand $1$ binds more strongly to $W$ than to
$U$, (ii) ligand $2$ binds more strongly to $W$ than ligand
$1$ does, and (iii) ligand-bound $U$ suppresses activation more strongly than
ligand-bound $W$ promotes activation. These are the precise
features that allow receptor $Q_{\rm UW}$ to perform the \gate{XOR}, as
outlined in detail in the previous section.

\begin{figure}
\begin{center}
\includegraphics[]{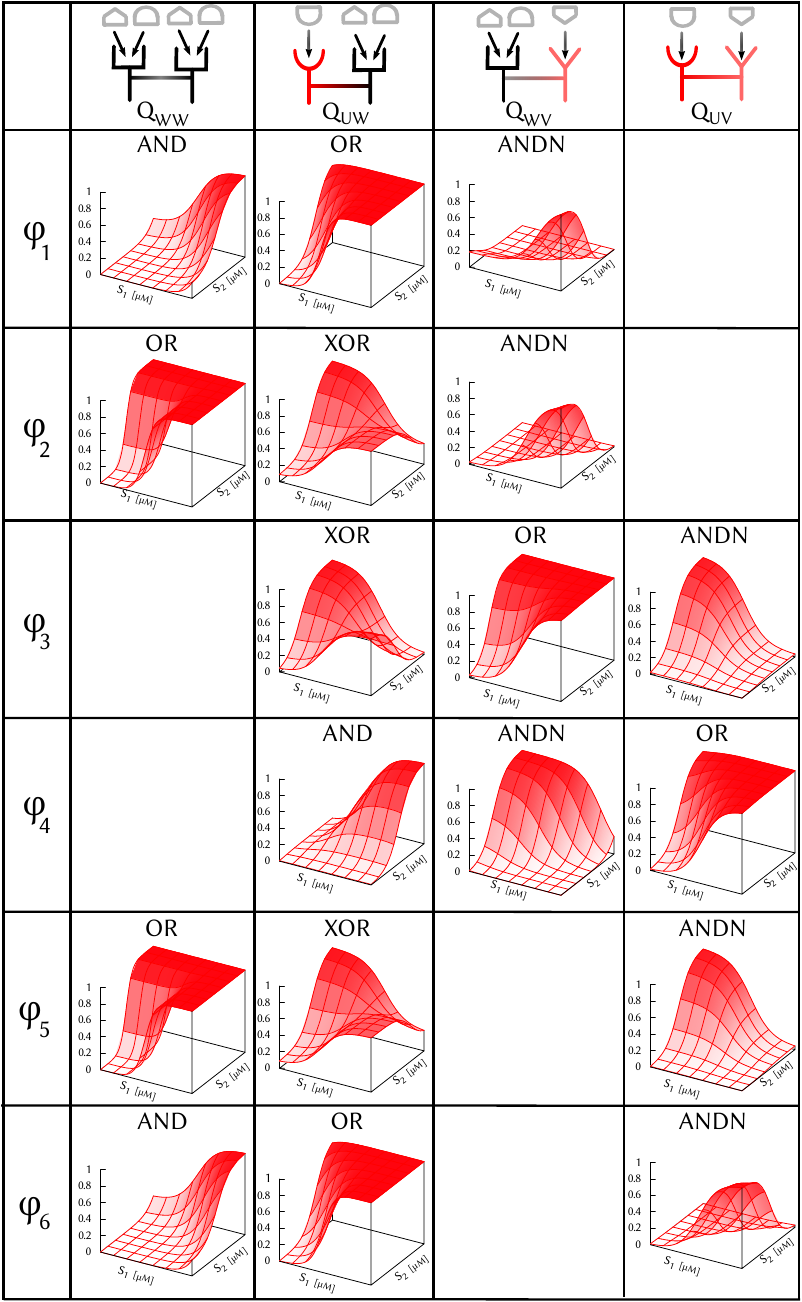}
\end{center}
\caption{\flabel{recombination} Functional versatility by recombination.
Given the three monomer types,
four functional units can be
formed by dimerization. Six parameter sets
$\varphi_k$ are shown at which three of the four dimers perform
functionally unique logic gates.}

\end{figure}

\begin{figure}
\begin{center}
\includegraphics[scale=2]{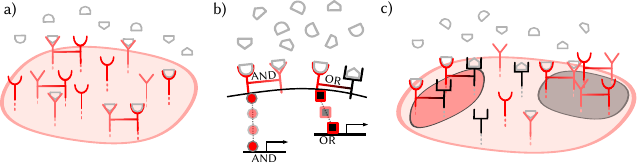}
\end{center}
\caption{\flabel{conflict} Several established mechanisms can minimize conflict
between dimers' logical functions.  (a) Selective expression of only two of the
three monomers allows the formation of only one functional dimer, while (b)
specificity of the downstream component or (c) spatial segregation of membrane
components allows multiple functional dimers to coexist without conflict.
}
\end{figure}

\section*{Discussion}

We have used a statistical mechanical model to investigate the versatility of
receptor function in two contexts: (i) the ability of a single
receptor to access logical functions by parameter variation, and (ii) the
ability --- at fixed parameters --- for a set of receptor monomers
to access logical functions by dimerizing. The first context is important on
evolutionary timescales, on which mutations and environmental
pressures act to change a cell's intrinsic biochemical parameters. The second
context is more critical at far shorter timescales, i.e.\
timescales characterizing the response of individual cells, during which gene
expression and covalent modification can potentially change cellular
function at the molecular level by favoring the dimerization of particular
receptors over others.

In the first context, we find that a single heterodimer (receptor $Q_{\rm
UW}$) can realize all possible logic gates by parameter variation.
Our analysis reveals that such complete functional freedom, while perhaps
surprising, is in fact quite intuitive for this receptor. In
particular, receptor $Q_{\rm UW}$ performs the most challenging function, the
\gate{XOR}, by exploiting an interference between the two
ligands (i.e.\ when both ligands are present, one outcompetes the other for
the activating binding pocket, ultimately causing suppression).
Such a nonmonotonic response requires competitive binding and asymmetric
activation biases, both of which are possible by
heterodimerization.

In the second context, we find that the simplest combination of monomers that
yields four functional dimers cannot in fact perform the four
unique logic gates at fixed parameters, an observation we explain by arguing
that the \gate{AND} and the \gate{XOR} are not mutually
accessible. Nonetheless, numerical search reveals that several distinct groups
of three unique gates are performable, a result that is
nontrivial given the high degree of overlap among dimers' parameter spaces. We
offer intuitive explanations for the emergence of these
groups, and further, we argue that these groups are exhaustive. The ability to
perform diverse functions with a limited set of simple
components is of critical importance to the question of how cells encode
reliable responses with limited resources.

Although we often think of logic functions as the fundamental units of
decision making, logic operations reduce the output space to a binary variable.
In principle the full input-output relation, which conveys much more
information, could be used to regulate downstream responses. Indeed, the
output of our statistical mechanical model is not restricted to Boolean logic,
but rather constitutes continuous response functions.  However, we
argue that the most simple form of transducing information on ligands is via
an input-output relation that approximates a binary response.
In fact, recent experiments have shown that the information transmission
capacity of a receptor is indeed approximately $1$ bit, which is
equivalent to a binary response \citep{Cheong2011}.

We have adopted a minimal model (the MWC model) to describe a minimal set of
components, and we have explored the functional capabilities
available under these conditions. We are further encouraged by the fact that
the MWC model has been shown to agree with experiments on
receptors \citep{Tu2008, Hansen2010, Skoge2006}. Nonetheless, several extensions
to the
model or the study itself are natural choices for further
exploration. First, the conformational spread (CS) model \citep{Changeux1967,
Bray2004} generalizes the MWC model, and thus it would allow
for more functional freedom in logic gate construction. However, it is always
a concern that generalizing one's model can reduce the
fraction of functional parameter space simply by increasing the total volume
of parameter space.
Second, it would be straightforward to introduce one or more additional
monomers when considering recombination. For example, introducing
an additional monomer that binds a single ligand might in fact admit parameter
sets at which all four unique logic gates are performed, at
the expense of increasing the number of individual components that the cell
must produce. The impact of such a finding, however, would be
reduced by the fact that more than four functional dimer combinations would be
possible. Third, it would also be straightforward to
consider more complex dimers (or higher oligomers), such as $Q_{W_1W_2}$, in
which each pocket binds both ligands competitively, but with
asymmetric parameters. Of course, such increasing complexity would only be
justified in the context of correspondingly detailed biological
examples.

It is well established that receptors are responsive to multiple ligands.
Recent experiments have indeed exploited this fact to
synthetically construct proteins that perform a limited set of logic gates
\citep{Lim2010a}. At the same time,
observations of oligomerization and protein modification on the
membrane
suggest that receptors can act as functional signaling units by
recombination. Indeed, experiments have shown that for many receptors, such as
ErbB and GPCR, monomers combine to form different dimers that
have different functionality \citep{Landau2008,Minneman2007, Citri2006a}.
We anticipate that this study will contribute to a predictive framework in
which experiments like these can be interpreted and extended.
The findings we report --- that a single receptor can function as any logic
gate and that a limited set of monomers can access diverse logic
gates by dimerizing --- speak to the large degree of functional control
available to cells at the level of individual receptor molecules.

\section*{Acknowledgements}

The authors would like to thank Jose Alvarado for a critical reading of the
manuscript. This work is part of the research programme of the
Foundation for Fundamental Research on Matter (FOM), which is financially
supported by the Netherlands Organization for Fundamental Research
(NWO).

\bibliography{protein_logic}
\newpage

\renewcommand{\theequation}{SI-\arabic{equation}}
\setcounter{equation}{0}
\renewcommand{\theequation}{SI-\arabic{equation}}
\renewcommand{\thefigure}{SI-\arabic{figure}}\renewcommand{\figurename}{Figure}

\renewcommand{\thetable}{SI-\arabic{table}}
\renewcommand{\thesection}{SI.~\arabic{section}}

\section{\label{app:num_opt}Optimization details}

\subsection*{Optimization parameters}

In \tref{app:table2} we provide bounds for our optimization parameters $K_i^j,
\omega_0$ and $\omega_{ii'}$. In this section we provide experimental support
for the chosen bounds.
\begin{itemize}
\item $\omega_{0}$\\
Two experimental studies report on explicit values of $\omega_0$. In the
quorum sensing bacterium {\em V. Harveyi} an $\omega_0$ value is
reported of $e^{-\Delta\epsilon/k_BT}=e^{3.2}\approx24$ \cite{Swem2008}.
Next, a
conformational spread model for the motor proteins in {\em E. coli} reported
for $\omega_0=e^{-E_A/k_BT}=e^{-0.66}=0.51$ \cite{Bai2010a}.
Moreover, two modeling studies have also suggested values for $\omega_0$. In
\cite{Duke2001} a value of $\omega_0=e^{-E_A/k_BT}=0.36$, while in
\cite{Motlagh2012} a large range for $\omega_0$ is suggested
$\omega_0\sim[e^{-10}-e^{10}]$.
\item $\omega_{ii'}$\\
For {\em E. coli} different experimental observations are reported, both for
the motor proteins and the receptors. The receptor coupling energy has been
reported to be around $0$ $k_BT$, leading to $\omega_{ii'}=1$
\cite{Hansen2010,
Tu2008}. For the motor protein, a value for $\omega_{ii'}=e^{E_{ii'}/k_BT}=62$
has been reported \cite{Bai2010a}. In a modeling study of the protein N-WASP
$\omega_{ii'}\sim 300$ has been used \cite{Prehoda2002a}. In the EGFR
receptor for monomers
$K_D\sim10^0\,[{\rm nM}]$, while for dimers $K_D\sim10^{-2}\,[{\rm nM}]$,
suggesting a positive coupling between two monomers ($\omega_{ii'} > 1$)
\cite{Jorissen2003}. In the same modeling studies as cited above,
estimates are $E_j=\ln \omega_{ii'} = 0$-$4$ $k_BT$ \cite{Duke2001}
and a large range $\omega_{ii'}\sim[e^{-10}-e^{15}]$ \cite{Motlagh2012}.
\item $K_i^j$\\
In {\em E. coli} dissociation constants for different ligands and activation
states for the Tar and Tsr receptors have been measured which vary between
$[10^{-2}-10^6]$ $[\rm mM]$ \cite{Hansen2010, Tu2008}. Experimental work on
the
receptors of the quorum sensing machinery in {\em V. Harveyi} where single
amino-acids are replaced have resulted in dissociation constant varying
between
$[10^0-10^5]$ $[\rm nM]$ \cite{Swem2008}. For EGFR different $K_D$'s are
reported for different dimer pairs, ranging from $10$ [pM] to $500$ [nM]
\cite{Jorissen2003, Landau2008}. Synthetic proteins are constructed with
varying
dissociation constants for different ligands, where the $K_D$ ranges from
$[1-1000]$ $[\rm \mu M]$ \cite{Dueber2003} or $[10^{-1}-10^5]$ $[\rm \mu M]$
\cite{Looger2003}. A mathematical model based upon a three-state
receptor with multiple ligands used $K_D$ values from $[10^{-9}-10^{-5}]$
$[\rm \mu M]$ for each
ligand and state of the system \cite{Leff1997}.
\end{itemize}

\begin{table}[h]
\begin{center}
\begin{tabular}{|l|c|}
\hline
\textbf{Model input} & \textbf{Range} \\
\hline
$\mol{S_1}, \mol{S_2}$ &
$\sqbr{10^{-2}-10^2}$ $\unit{\mu M}$ \\
\hline\hline
\textbf{Model parameter} & \textbf{Bounds} \\
\hline
$K^j_{i,k}$ &  $\sqbr{10^{-3}-10^4}$ $\unit{\mu M}$
\\
\hline
$\omega_0$ & $\sqbr{10^{-3}-10^3}$ \\
\hline
$\omega_{ii'}$ & $\sqbr{10^{-2}-10^2}$ \\
\hline\hline
\textbf{Optimization parameter} & \textbf{Value} \\
\hline
$\Delta$ & 0.3\\
\hline
$N$ & 4\\
\hline
$R$ & $50000$\\
\hline
Steps & $1000$\\
\hline
\end{tabular}
\end{center}
\caption{\tlabel{app:table2} Overview of parameters.  During optimization,
model parameters are initialized and constrained within the
indicated bounds.
}
\end{table}

\subsection*{Multiple receptors}
In the case where $M$ different receptors (e.g. $Q_{\rm UV},Q_{\rm WV},Q_{\rm
UW}$) are combined to act in different combinations as
unique logic gates, the optimization algorithm follows a specific order in the
optimization.
A straightforward extension of the model for a single receptor is the
optimization of three (or four) gates simultaneously, and
taking as fitness ${\cal F}$ the summed fitness of every gate $F_m$:
\begin{align}
{\cal F}=\sum_{m=1}^M F_m.
\end{align}
However, this optimization is not capable of optimizing all the gates
independently. Instead, the algorithm optimizes either one (or two)
gates, but then cannot optimize the third gate. To optimize the third gate,
the already optimized gates decrease (temporarily) in fitness.
This decrease is larger than the increase in fitness for the third gate and
the algorithm finds suboptimal peaks in this rugged fitness
landscape.

Instead of optimizing all gates simultaneously, we optimize gates in order.
For the homodimer construction ($Q_{\rm WW}, Q_{\rm UW},
Q_{\rm WV}$), we first optimize $Q_{\rm WW}$, then $Q_{\rm UW}$, where we only
change the parameters of $U$, and then $Q_{\rm WV}$, only
changing $V$. The achieved results greatly outperform the results where we
optimize all three gates simultaneously.

For the heterodimer construction ($Q_{\rm UW}, Q_{\rm WV}, Q_{\rm UV}$), we
again start by optimizing gate $Q_{\rm UW}$, then the two gates
$Q_{\rm UW}$ and $Q_{\rm WV}$ simultaneously, and finally $Q_{\rm UW},$
$Q_{\rm WV}$, and $Q_{\rm UV}$. Again this procedure gives much
better results
than simultaneous optimization of all three gates.

For the construction with $Q_{\rm WW}, Q_{\rm UW}, Q_{\rm UV}$), we start by
optimizing gate $Q_{\rm WW}$, then the two gates
$Q_{\rm WW}$ and $Q_{\rm UW}$ simultaneously, and finally $Q_{\rm WW},$
$Q_{\rm UW}$, and $Q_{\rm UV}$.

\section{\label{app:proof}Formal proof that receptor $Q_{\rm UW}$ can perform
an \gate{XOR}}
The probability to be active $p^A\br{\mol{S_1},\mol{S_2}}$ in an \gate{XOR} is
a nonmonotonic function of $\mol{S_1}$ and $\mol{S_2}$ simultaneously.  More
specifically, for constant $\mol{S_2}=\mol{S^c_2}$,
$p^A\br{\mol{S_1},\mol{S^c_2}}$ is either monotonically increasing or
decreasing, depending on the value $\mol{S^c_2}$: for small $\mol{S^c_2}$,
$p^A$
is monotonically increasing, while for large $\mol{S^c_2}$, $p^A$ is
monotonically decreasing.

A positive derivative $\partial p^A/\partial [S_2]$ reflects a monotonically
increasing function, while a negative derivative reflects a monotonically
decreasing function. Therefore, in an \gate{XOR}, the derivative of the
probability with respect to $\mol{S_2}$ at constant $\mol{S^c_1}$ should
change
sign as function of $\mol{S^c_1}$. Again due to symmetry, the derivative of
the
probability with respect to $\mol{S_1}$ at constant $\mol{S^c_2}$ should
change
sign as function of $\mol{S^c_2}$. We will prove that the \gate{XOR} is
possible
for the $Q_{\rm UW}$ receptor even with $\omega_{11}=\omega_{21}=1$. Recalling
Eq.~1, the derivative can be written $\partial p^A/\partial [S_2] =
f/(Z^A+Z^I)^2$, where the numerator
\begin{align}
\elabel{app:sign}
f\br{\mol{S^c_1},\mol{S_2}}=\pdiff{Z^A}{\mol{S_2}}Z^I-Z^A\pdiff{Z^I}{\mol{S_2}}
\end{align}
alone determines the sign.
We therefore must show that $f$ changes sign as function of
$\mol{S^c_1}$.
Specifically, the \gate{XOR} requires
\begin{align}
\elabel{app:gateXOR1a}
f&>0{\text{ for }} \mol{S^c_1}<[S^c_1]^*,\\
\elabel{app:gateXOR1b}
f&<0{\text{ for }} \mol{S^c_1}>[S^c_1]^*,
\end{align}
for some $[S^c_1]^*$ and all $[S_2]$.

The partition functions $Z^A$ and $Z^I$ for the $Q_{\rm UW}$
receptor are given by Eqs.~9-10 in the main text.
Performing the derivatives in \eref{app:sign} reveals that $f$ is a third
order
polynomial in $\mol{S_1^c}$ in which all dependence on $[S_2]$
drops out.  Only one root is potentially positive:

\begin{align}
\elabel{app:root1}
\mol{S^c_1}^*=\frac{K_{1,W}^IK_{1,W}^A\br{K_{2}^I-K_{2}^A}}{K_{1,W}^IK_{2}
^A-K_{
1,W}^AK_{2}^I}.
\end{align}
To satisfy \erefs{app:gateXOR1a}-\ref{eqn:app:gateXOR1b}, we require that the
zeroth order term (the intercept) is positive and that the leading order term
is negative; enforcing these conditions yields
\begin{align}
K^I_{2}-K^A_{2}>0 \elabel{app:constr1},\\
K^I_{1,W}K^A_{2}-K^A_{1,W}K^I_{2}>0 \elabel{app:constr2},
\end{align}
which are in fact the precise conditions that maintain positivity of the root
(\eref{app:root1}).  Parameters that satisfy these conditions enable the sign
of
$\partial p^A/\partial [S_2]$ to depend on constant $[S_1^c]$, which is one of
the two conditions necessary to perform the \gate{XOR}.  Notably,
\eref{app:constr1} directly shows that the binding of ligand $2$ to the $W$
monomer in $Q_{\rm UW}$ in the active state is less likely than binding in the
inactive state.

The second requirement is that the sign of $\partial p^A/\partial
[S_1]$ depends on constant $[S_2^c]$.
Specifically,
as in \erefs{app:gateXOR1a}-\ref{eqn:app:gateXOR1b}, the \gate{XOR} requires
that the numerator $g(\mol{S_1}, \mol{S^c_2})$ of the derivative satisfies
\begin{align}
\elabel{app:gateXOR2a}
g&>0{\text{ for }} \mol{S^c_2}<[S^c_2]^*,\\
\elabel{app:gateXOR2b}
g&<0{\text{ for }} \mol{S^c_2}>[S^c_2]^*.
\end{align}
for some $[S^c_2]^*$ and all
$[S_1]$. Performing the derivative reveals that $g$ is a second order
polynomial
whose coefficients depend on $[S_1]$. To satisfy
\erefs{app:gateXOR2a}-\ref{eqn:app:gateXOR2b}, we again require that the
intercept is positive and that the leading order term is negative; enforcing
these conditions yields
\begin{align}
\elabel{app:constr3}
h\br{[S_1],K^I_{1,U},K^A_{1,U},K^I_{1,W},K^A_{1,W}}&>0,\\
\elabel{app:constr4}
K^I_{1,U}-K^A_{1,U}&<0,
\end{align}
where the function $h$ results straightforwardly from the derivative but is
unwieldy, such that we do not reproduce it here. The roots of the polynomial
$[S_2^c]^*$ are similarly unwieldy, but noting positivity requirements
($[S^c_2]^*>0$, $K^I_{1,W}>0$, $K^I_{1,U}>0$, $K^I_{2}>0$), parameter regimes
can be derived that satisfy both \erefs{app:constr1}-\ref{eqn:app:constr2} and
\erefs{app:constr3}-\ref{eqn:app:constr4} simultaneously.
As an example we present one possible regime here:
\begin{align}
\elabel{res3}
\frac{K^A_{1,W}}{K^I_{1,W}}&<\frac{K^A_{2}}{K^I_{2}}<1,\\
\elabel{res1}
0<\frac{K^A_{1,W}}{K^I_{1,W}}&\le\frac{K^I_{1,U}}{K^I_{1,U}+K^I_{1,W}},\\
\elabel{res2}
K^A_{1,W}+K^A_{1,U}&>K^I_{1,U}+K^I_{1,W}.
\end{align}
\eref{res3} states that the $W$ monomer is activated by $\mol{S_1}$ and
$\mol{S_2}$, and that activation by $\mol{S_1}$ is stronger than activation by
$\mol{S_2}$. Note that for small concentrations of either $\mol{S_1}$ or
$\mol{S_2}$, $W$ is inactive. The two more interesting constraints are in
\eref{res1} and \eref{res2}. \eref{res2} states that $\mol{S_1}$ bound to
monomer $U$ deactivates the receptor ($K^A_{1,U}>K^I_{1,U}$), since, from
\eref{res3}, we have seen that $K^A_{1,W}<K^I_{1,W}$. More importantly the
deactivation of $U$ by binding $\mol{S_1}$ is stronger than the activation of
$W$ by $\mol{S_1}$, and following \eref{res3}, it is thus also stronger than
activation of $W$ by $\mol{S_2}$. This is precisely the interference
interaction
as described in the main text. The last constraint, \eref{res1}, provides the
required binding strength of $\mol{S_1}$ to $U$ and $W$ to satisfy all
constraints. In the main text we argue that $W$ should preferably bind
$\mol{S_2}$, such that in the presence of both ligand $\mol{S_2}$ binds to $W$
and $\mol{S_1}$ binds to $U$ with as result that $Q_{UW}$ is inactive.

Here we have shown that the $Q_{\rm UW}$ receptor is capable of the
nonmonotonic
derivatives required by the \gate{XOR}. This capability is necessary but not
sufficient to perform the gate, as an ideal logic gate requires that the
output
be maximally high and low at the appropriate input values.  Our numerical
results, however, indeed confirm that the $Q_{\rm UW}$ receptor can perform
the
\gate{XOR}.

\section{\label{app:par_sens}Parameter sensitivity}

In this section we discuss the sensitivity to parameter variation of the
results
at the six parameter sets $\varphi_k$ shown in Fig.~3.  Robustness against
parameter fluctuations generally is considered an important quality of
biochemical systems, due to stochastic nature of intra- and extracellular
processes. If the observed logic gates only function within a very narrow
parameter regime, this could lead to unreliable functioning.

Parameters are varied according to
\begin{align}
\varphi^z_{\rm new}=\varphi^z_{\rm old}\br{1+n^z}
\end{align}
where $n^z$ is the $z$th component of a uniformly distributed random vector
$\bf{n}$ with norm $\abs{\bf{n}}=\eta$.  Under this implementation, $\eta$
sets
the average (root mean square) factor by which each parameter changes via
$\avg{\delta\varphi^z/\varphi^z} = \eta/\sqrt{Z}$, where $Z$ is the number of
parameters.
We sample $10^6$ different vectors $\bf{n}$.

Sensitivity is measured by computing the fraction of new parameter sets for
which, for each individual gate $m$, the relative change in fitness is less
than
a factor $\lambda$:
\begin{align}
\frac{\abs{F_m^{\rm new}-F_m^{\rm old}}}{F_m^{\rm old}}<\lambda \quad \forall
\, m.
\end{align}
Figure \ref{fig:robust} reveals that for all $\varphi_k$, most random
perturbations in which each parameter changes by an average of
$\avg{\delta\varphi^z/\varphi^z} \sim 20\%$ change the fitness of none of the
three logic gates by more than $\lambda = 10\%$.

\begin{figure}[ht]
\begin{center}
\includegraphics[]{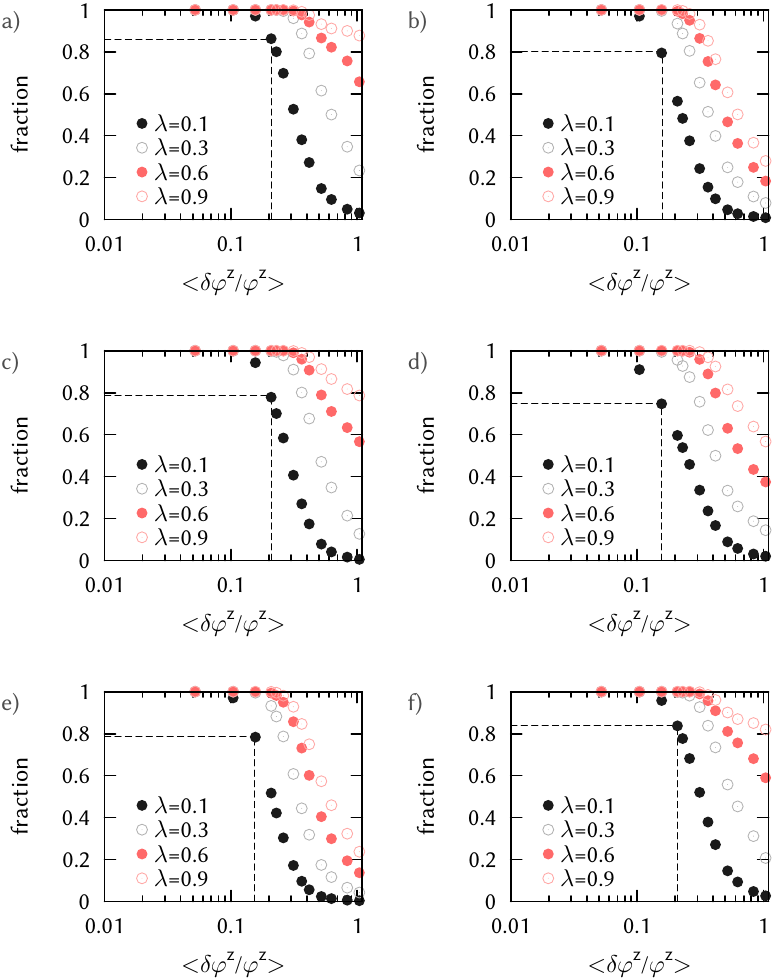}
\end{center}
\caption{\flabel{robust} Robustness to parameter variation for the six parameter
sets at which dimers
can form three unique logic
gates (Fig.~3): (a) $\varphi_1$, (b) $\varphi_2$, (c) $\varphi_3$, (d)
$\varphi_4$, (e) $\varphi_5$, and (f) $\varphi_6$. An
increase in
$\avg{\delta\varphi^z/\varphi^z}$ reflects a larger range of parameter
fluctuations and an increase in $\lambda$ reflects a
loosening of the robustness constraint. The dashed black lines indicate that a
significant fraction of random perturbations in which each
parameter changes by an average of $\avg{\delta\varphi^z/\varphi^z} \sim 20\%$
change the fitness of none of the
three logic gates by more than $\lambda = 10\%$.
}
\end{figure}

\section{\label{app:elimination} Further intuition behind functions accessible
by recombination}

In the main text, we provide the intuition behind how the first two groups of
logic gates in Fig.~3 are performed by the
corresponding
receptors. Here, we provide similar intuition for the last four groups.  Then,
we argue why the six groups observed in Fig.~3 (and their
counterparts obtained upon ligand exchange) are the
{\it only} groups of three unique logic gates that one expects to observe
under this model.

\subsection*{Parameter sets not discussed in the main text}

Parameter set $\varphi_3$ (Fig.~3, third row) is similar to set $\varphi_2$
(see discussion of $\varphi_2$ in main text).  In particular,
receptor $Q_{\rm UW}$ performs the
\gate{XOR} in the same way.  The difference between $\varphi_3$ and
$\varphi_2$ is that ligand-bound $V$ promotes activation instead of
suppressing activation.  Since ligand-bound $W$ also promotes activation, this
feature allows receptor $Q_{\rm WV}$ to perform the
\gate{OR}.  However, ligand-bound $U$ suppresses activation more strongly than
ligand-bound $V$ promotes activation.  This feature allows
receptor $Q_{\rm UV}$ to perform the \tt{ANDN}$_{S_1}$ gate, since only in the
presence of ligand $2$ and not $1$ will activation be
promoted via $V$ and not suppressed via $U$.

Parameter set $\varphi_4$ (Fig.~3, fourth row) corresponds to a case where
ligand-bound $U$ and ligand-bound $V$ both promote
activation. This feature is sufficient for receptor $Q_{\rm UV}$ to perform
the \gate{OR}.  Furthermore, ligand $2$ binds more strongly to
$V$ than to $W$, and ligand-bound $W$ suppresses activation more strongly than
ligand-bound $V$ promotes activation.  These features allow
receptor $Q_{\rm WV}$ to perform the \tt{ANDN}$_{S_1}$ gate, since only in the
presence of ligand $2$ and not $1$ will activation be
promoted via $V$ and not suppressed via $W$.  Finally, (i) ligand $1$ binds
more strongly to $W$ than to $U$, (ii) ligand $2$ binds more
strongly to $W$ than ligand $1$ does, and (iii) ligand-bound $U$ promotes
activation more strongly than ligand-bound $W$ suppresses
activation.  These three features allow receptor $Q_{\rm UW}$ to perform the
\gate{AND}: when ligand $1$ is present alone, feature (i)
results in suppression via $W$; when ligand $2$ is present alone, it only
binds to $W$, resulting in suppression; and when both ligands are
present, features (ii) and (iii) cause ligand $2$ to bind to $W$, forcing
ligand $1$ to bind to $U$ and thus activating the receptor.

Parameter set $\varphi_5$ (Fig.~3, fifth row) is once again similar to set
$\varphi_2$. In particular, receptors $Q_{\rm WW}$ and $Q_{\rm
UW}$ perform the \gate{OR} and the \gate{XOR} in the same way, respectively.
Additionally, ligand $1$ suppresses activation via $U$ more
strongly than ligand $2$ promotes activation via $V$.  This feature allows
receptor $Q_{\rm UV}$ to perform the \tt{ANDN}$_{S_1}$ gate,
since only in the presence of ligand $2$ and not $1$ will the receptor be
active.

Parameter set $\varphi_6$ (Fig.~3, sixth row) is similar to set $\varphi_1$
(see discussion of $\varphi_1$ in main text).  In particular,
receptors $Q_{\rm WW}$ and $Q_{\rm UW}$ perform the \gate{AND} and the
\gate{OR} in the same way, respectively.
Additionally, ligand $2$ suppresses activation via $V$ more strongly than
ligand	$1$ promotes activation via $U$.  This feature allows
receptor $Q_{\rm UV}$ to perform the \tt{ANDN}$_{S_2}$ gate, since only in the
presence of ligand $1$ and not $2$ will the receptor be
active.

\subsection*{Figure~3 is exhaustive}

Here, we argue why the groups observed in Fig.~3 (and their counterparts
obtained upon ligand exchange) are the
only groups of three unique logic gates that one expects to observe under this
model.  The overall logic is presented first, with the
arguments subsequently given in subsections.

There are $4$ ways to choose a group of three from the four functional dimers
$Q_{\rm WW}$, $Q_{\rm UW}$, $Q_{\rm WV}$, and $Q_{\rm UV}$ to
perform the three unique logic gates: $\{Q_{\rm WW}, Q_{\rm UW}, Q_{\rm
WV}\}$, $\{Q_{\rm UW}, Q_{\rm WV}, Q_{\rm UV}\}$, $\{Q_{\rm WW},
Q_{\rm UW}, Q_{\rm UV}\}$, and $\{Q_{\rm WW}, Q_{\rm WV}, Q_{\rm UV}\}$.  The
last two groups are symmetric upon ligand exchange; we
therefore
consider only the first three groups.

The first group is $\{Q_{\rm WW}, Q_{\rm UW}, Q_{\rm WV}\}$.  As shown in the
main text, receptor $Q_{\rm WW}$ is capable of performing an
\gate{AND}, an \gate{OR}, or an \gate{ANDN}, but not an \gate{XOR} (Fig.~2).
If receptor $Q_{\rm WW}$ performs an \gate{AND}, receptor $Q_{\rm UW}$ can
perform an \gate{ANDN} or an \gate{OR}, but not an \gate{XOR}
(Argument 1).  Receptor $Q_{\rm WV}$ then performs the \gate{OR} or the
\gate{ANDN}, respectively (it also cannot perform an \gate{XOR} by
the same argument).  These two possibilities are represented by parameter set
$\varphi_1$ (Fig.~3) and its counterpart upon
ligand exchange.
If receptor $Q_{\rm WW}$ performs an \gate{OR}, receptor $Q_{\rm UW}$ can
perform an \gate{ANDN} or an \gate{XOR}, but not an \gate{AND}
(Argument 2).  Receptor $Q_{\rm WV}$ then performs the \gate{XOR} or the
\gate{ANDN}, respectively (it also cannot perform an \gate{AND} by
the same argument).  These two possibilities are represented by parameter set
$\varphi_2$ (Fig.~3) and its counterpart upon
ligand exchange.
If receptor $Q_{\rm WW}$ performs an \gate{ANDN}, three unique gates cannot be
performed (Argument 3).  Therefore, this group is
exhaustively represented by $\varphi_1$ and $\varphi_2$.

The second group is $\{Q_{\rm UW}, Q_{\rm WV}, Q_{\rm UV}\}$.  As shown in the
main text, receptor $Q_{\rm UV}$ is capable of performing an
\gate{ANDN}, an \gate{OR}, or an \gate{AND}, but not an \gate{XOR} (Fig.~2).
If receptor $Q_{\rm UV}$ performs an \gate{ANDN}, receptor $Q_{\rm UW}$ can
perform an \gate{XOR} or an \gate{OR}, but not an \gate{AND}
(Argument 4).  Receptor $Q_{\rm WV}$ then performs the \gate{OR} or the
\gate{XOR}, respectively (it also cannot perform an \gate{AND} by
the same argument).  These two possibilities are represented by parameter set
$\varphi_3$ (Fig.~3) and its counterpart upon
ligand exchange.
If receptor $Q_{\rm UV}$ performs an \gate{OR}, receptor $Q_{\rm UW}$ can
perform an \gate{AND} or an \gate{ANDN}, but not an \gate{XOR}
(Argument 5).  Receptor $Q_{\rm WV}$ then performs the \gate{ANDN} or the
\gate{AND}, respectively (it also cannot perform an \gate{XOR} by
the same argument).  These two possibilities are represented by parameter set
$\varphi_4$ (Fig.~3) and its counterpart upon
ligand exchange.
If receptor $Q_{\rm UV}$ performs an \gate{AND}, three unique gates cannot be
performed (Argument 6).  Therefore, this group is exhaustively
represented by $\varphi_3$ and $\varphi_4$.

The third group is $\{Q_{\rm WW}, Q_{\rm UW}, Q_{\rm UV}\}$. We note that this
group is different from
the first two groups, since it does not contain the two receptors $Q_{\rm UW}$
and $Q_{\rm UV}$ which are symmetric upon ligand exchange.
As shown in the main text, receptor $Q_{\rm WW}$ is capable of performing an
\gate{AND}, an \gate{OR}, or an \gate{ANDN}, but not an
\gate{XOR} (Fig.~2). If receptor $Q_{\rm WW}$ performs an \gate{AND}, receptor
$Q_{\rm UW}$ can perform an \gate{ANDN}
or an \gate{OR}, but not an \gate{XOR} (Argument 1). If receptor $Q_{\rm UW}$
performs an \gate{ANDN},
receptor $Q_{\rm UV}$ cannot perform an \gate{OR} (Argument 7); since receptor
$Q_{\rm UV}$ also cannot perform an \gate{XOR}
(Fig.~2), three unique gates cannot be performed.  Therefore, receptor $Q_{\rm
UW}$ must perform an \gate{OR}, leaving receptor
$Q_{\rm UV}$ to perform an \gate{ANDN}.  This possibility is represented by
parameter set $\varphi_6$ (Fig.~3).
If receptor $Q_{\rm WW}$ performs an \gate{OR}, receptor $Q_{\rm UW}$ can
perform an \gate{ANDN}
or an \gate{XOR}, but not an \gate{AND} (Argument 2). If receptor $Q_{\rm UW}$
performs as an \gate{ANDN},
receptor $Q_{\rm UV}$ cannot perform an \gate{AND} (Argument 8); since
receptor	$Q_{\rm UV}$ also cannot perform an \gate{XOR}, three unique
gates cannot be performed.  Therefore, receptor $Q_{\rm UW}$ must perform an
\gate{XOR}, leaving receptor $Q_{\rm UV}$ to perform an
\gate{ANDN}.  This possibility is represented by parameter set $\varphi_6$
(Fig.~3).
If receptor $Q_{\rm WW}$ performs as an \gate{ANDN}, three unique gates can
not be performed (Argument 9).
Therefore, this group is exhaustively represented by $\varphi_5$ and
$\varphi_6$.

This completes the logic arguing that the groups observed in Fig.~3 are
exhaustive.

\subsubsection*{Argument 1}

If receptor $Q_{\rm WW}$ performs an \gate{AND}, ligand $2$ alone does not
promote activation when binding to monomer $W$.  Therefore,
because ligand $2$ does not bind to monomer $U$, the receptor $Q_{\rm UW}$ is
always inactive if ligand $2$ is present alone.  This behavior
is inconsistent with the logic of the \gate{XOR}.

\subsubsection*{Argument 2}

If receptor $Q_{\rm WW}$ performs an \gate{OR}, ligand $2$ alone promotes
activation when binding to monomer $W$.  Therefore, because ligand
$2$ does not bind to monomer $U$, the receptor $Q_{\rm UW}$ is always active
if ligand $2$ is present alone.  This behavior is inconsistent
with the logic of the \gate{AND}.

\subsubsection*{Argument 3}

If receptor $Q_{\rm WW}$ performs an \gate{ANDN}, receptors $Q_{\rm UW}$ and
$Q_{\rm WV}$ can each perform neither an \gate{XOR} nor an
\gate{AND}, thereby preventing the group $\{Q_{\rm WW}, Q_{\rm UW}, Q_{\rm
WV}\}$ from performing three unique gates.
The reason is straightforward: if receptor $Q_{\rm WW}$ performs an
\gate{ANDN}, one ligand must suppress activation via $W$ while the other
ligand promotes activation via $W$.  This feature immediately excludes the
\gate{XOR} since, as described in the main text, an \gate{XOR}
requires both ligands to promote activation via $W$.  This feature also
excludes an \gate{AND} since, as also described in the main text, an
\gate{AND} requires either that activation via $W$ is promoted only weakly or
that both ligands suppress activation via $W$.  In the first
case, activation of receptor $Q_{\rm UW}$ (or $Q_{\rm WV}$) is only achieved
cooperatively when both ligands are present.  In the second
case, activation is achieved with both ligands present via $U$ (or $V$) due to
an interference effect similar to that underlying the
\gate{XOR} (see discussion of parameter set $\varphi_4$ above).

\subsubsection*{Argument 4}

If receptor $Q_{\rm UV}$ performs an \gate{ANDN$_{S_1}$}, $Q_{\rm UW}$ cannot
function as an \gate{AND}. To function as an \gate{AND} (see
Eq.~9), this requires that $\omega_0 K^A_{1,U}K^A_{2}\ll K^I_{1,U}K^I_{2}$,
while $\omega_0K^A_{1,U}\gg
K^I_{1,U}$ and $\omega_0K^A_{2}\gg K^I_{2}$. This
conditions cannot be satisfied simultaneously.

\subsubsection*{Argument 5}

If receptor $Q_{\rm UV}$ performs an \gate{OR}, ligand $1$ activates the
receptor via $U$.  However, for receptor $Q_{\rm UW}$ to perform
the \gate{XOR}, ligand $1$ must suppress activation via $U$, as described in
the main text.

\subsubsection*{Argument 6}

If $Q_{\rm UV}$ functions as an \gate{AND} $U$ is activated by $S_1$ and $V$
is activated by $S_2$, but both activation biases alone are
insufficient to activate the receptor. This excludes the formation of a
\gate{XOR} for either the $Q_{\rm UW}$ or $Q_{\rm WV}$. As we have
discussed in the previous section, the \gate{XOR} is obtained by the
deactivation of $U$ ($V$) by ligand $S_1$ ($S_2$).
However, it is possible that $Q_{\rm UW}$ is a \gate{OR}, while $Q_{\rm WV}$
is a \gate{ANDN}. The $Q_{\rm UW}$-\gate{OR} requires that
$W$ is activated by $S_2$ and $S_1$, since monomer $U$ is not active in the
presence of $S_1$.  The $Q_{\rm WV}$-\gate{ANDN} requires that
$W$ is strongly deactivated by $S_1$. These two conditions on $W$ are mutually
exclusive.

\subsubsection*{Argument 7}
If receptors $Q_{\rm WW}$ and $Q_{\rm UW}$ perform an \gate{AND} and an
\gate{ANDN}, respectively, the \gate{ANDN} must be
\tt{ANDN}$_{S_1}$, not \tt{ANDN}$_{S_2}$.  The reason is that the \gate{AND}
requires ligand $2$ to promote activation via $W$, while the
\tt{ANDN}$_{S_2}$ gate requires ligand $2$ to suppress activation via $W$.
Then, if $Q_{\rm UW}$ indeed performs the \tt{ANDN}$_{S_1}$
gate, receptor $Q_{\rm UV}$ cannot perform an \gate{OR}.  The reason is that
the \tt{AND} and \tt{ANDN}$_{S_1}$ gates require ligand $1$ to
suppress activation via $U$ and not via $W$, while the \gate{OR} requires
ligand $1$ to promote activation via $U$.

\subsubsection*{Argument 8}
If receptors $Q_{\rm WW}$ and $Q_{\rm UW}$ perform an \gate{OR} and an
\gate{ANDN}, respectively, the \gate{ANDN} must be \tt{ANDN}$_{S_1}$,
not \tt{ANDN}$_{S_2}$.  The reason is that the \gate{OR} requires ligand $2$
to promote activation via $W$, while the \tt{ANDN}$_{S_2}$ gate
requires ligand $2$ to suppress activation via $W$.  Then, if $Q_{\rm UW}$
indeed performs the \tt{ANDN}$_{S_1}$ gate, receptor $Q_{\rm UV}$
cannot perform an \gate{AND}.  The reason is that the \tt{OR} and
\tt{ANDN}$_{S_1}$ gates require ligand $1$ to suppress activation via $U$
and not via $W$, while the \gate{AND} requires ligand $1$ to promote
activation via $U$.

\subsubsection*{Argument 9}
If receptor $Q_{\rm WW}$ performs an \gate{ANDN}, receptor $Q_{\rm UW}$ cannot
perform a \gate{XOR} gate, since this requires that
both ligands activate $W$.
If receptor $Q_{\rm WW}$ performs an \tt{ANDN}$_{S_1}$ gate, receptor $Q_{\rm
UW}$ cannot perform an \gate{AND}, since $Q_{\rm UW}$ is
active if only ligand $2$ is present.
If receptor $Q_{\rm WW}$ performs an \tt{ANDN}$_{S_1}$ gate, receptor $Q_{\rm
UW}$ can perform an \gate{OR} if ligand $1$ activates $U$
more strongly than it deactivates $W$. However, receptor $Q_{\rm UV}$ is then
always active if ligand $1$ is present, and this is
inconsistent
with the logic of the \gate{AND}.
If receptor $Q_{\rm WW}$ performs an \tt{ANDN}$_{S_2}$ gate, receptor $Q_{\rm
UW}$ cannot perform an \gate{AND}, since $Q_{\rm UW}$ is
active if only ligand $1$ is present (ligand $1$ activates $U$) or $Q_{\rm
UW}$ is never active (ligand $1$ deactivates $U$ more strongly
than
it activates $W$).
If receptor $Q_{\rm WW}$ performs an \tt{ANDN}$_{S_2}$ gate, receptor $Q_{\rm
UW}$ can perform an \gate{OR}, if (i) ligand $1$
activates $U$ and (ii) in the presence of small ligand $1$ and an abundance of
ligand $2$ the receptor $Q_{\rm UW}$ is active.
However, receptor $Q_{\rm UV}$ is then always active if ligand $1$ is present
and this is inconsistent with the logic of the \gate{AND}.

\end{document}